\documentclass[prd,amsmath,floats,amssymb, floatfix,superscriptaddress,nofootinbib,onecolumn]{revtex4} 

\usepackage[plainpages=false, colorlinks=true, anchorcolor=blue, linkcolor=blue, citecolor=blue, bookmarks=false]{hyperref}
\usepackage[T1]{fontenc}
\usepackage{graphicx}
\usepackage{dcolumn}
\usepackage{bm}
\usepackage{longtable}
\usepackage{adjustbox}
\newcommand{\rthis}[1]{\textcolor{black}{#1}}
\usepackage{float}
\usepackage[caption=false]{subfig}
\usepackage[paperwidth=210mm,paperheight=297mm,centering,hmargin=2cm,vmargin=2.5cm]{geometry}

\begin{document}
\include{notations}
\preprint{APS/123-QED}

\title{A Stacked Analysis of GeV Gamma-Ray Emission from SPT-SZ Galaxy Clusters with 16 Years of Fermi-LAT Data}

\author{Siddhant Manna}
 \altaffiliation{Email:ph22resch11006@iith.ac.in}
\author{Shantanu Desai}
 \altaffiliation{Email:shntn05@gmail.com}
\affiliation{
 Department of Physics, IIT Hyderabad Kandi, Telangana 502284,  India}





\begin{abstract}
We report a statistically significant detection of cumulative $\gamma$-ray emission from a stacked sample of SPT-SZ selected galaxy clusters using 16.4 years of data from the Fermi Gamma-Ray Space Telescope's Large Area Telescope (LAT). By analyzing a population of clusters with individual Test Statistic (TS) values $<$ 9.0, we identify a robust cumulative signal with a TS of 75.2, which corresponds to approximately $8.4\sigma$ significance. In contrast, clusters with TS $<$ 4.0 yield a weaker cumulative signal of TS equal to  9.65 (2.65$\sigma$), consistent with background fluctuations. The derived $\gamma$-ray spectrum is well characterized by a power law model with a best-fit spectral index of $-2.59 \pm 0.20$ and an integrated flux of $1.67^{+1.35}_{-1.07} \times 10^{-11}$ ph cm$^{-2}$ s$^{-1}$. \rthis{The  high TS values in the full sample are likely driven by AGN-dominated clusters. However,  when clusters with TS between 4 and 9 are excluded, the remaining low-TS population shows a stacked signal consistent with hadronic emission.}
\end{abstract}


\maketitle
\section{\label{sec:level1}Introduction\protect}
Galaxy clusters, the most massive gravitationally bound systems in the universe, comprise hundreds to thousands of galaxies embedded in a hot intracluster plasma and enveloped by massive dark matter halos~\cite{Davis1985}. These systems serve as exceptional laboratories for testing cosmology~\cite{Allen2011,Vikhlinin2014,Kravtsov2012}, and fundamental physics~\cite{Desai2018,Bora2021,Bora2021b,Bora2022,Boehringer2016,Bora2023,Mendoncca2021}. They have been extensively observed across the electromagnetic spectrum from radio frequencies to high-energy X-rays~\cite{Feretti2012,Wik2014,Birkinshaw1999,Bulbul2019,Pfrommer2008}.
Despite this wealth of multi-wavelength data, the detection of $\gamma$-ray emission ($\geq 100$~MeV) from clusters remains an unsettled issue. 

The current state of the art gamma-ray telescope is the  Large Area Telescope (LAT) aboard the Fermi Gamma-ray Space Telescope. The LAT  is a pair-conversion telescope optimized to detect high-energy photons in the energy range of 20~MeV to over 300~GeV. Launched on June~11, 2008, the LAT achieves a wide field of view (FoV) that covers $\sim$ 20\% of the sky instantly, allowing full-sky surveys every three hours~\cite{Atwood2009}. 
Its unprecedented sensitivity has revolutionized $\gamma$-ray astronomy unveiling a plethora of new sources~\cite{Ackermann2012}. 

To resolve the conundrum related to gamma-ray emission from clusters,  Fermi-LAT has been extensively used for a large number of  searches for  gamma-ray emissions from galaxy clusters since its launch in 2008, for signals of different types such as  diffuse broadband~\cite{Abdo2010,Arlen2012,Han2012,Prokhorov2014,Ackermann2014,Ackermann2015,Ackermann2016,Branchini2017,Dutson2013,Griffin2014,Prokhorov2014b,Shirasaki2020,Zandanel2014} and line emissions ~\cite{Fan2024,Shen2021,Anderson2016,Liang2016}. A comprehensive review of these efforts is summarized in~\cite{Manna2024}, which also summarizes some of  the proposed mechanisms for the production of gamma-rays in clusters.

While the Fermi-LAT collaboration has reported stringent upper limits on $\gamma$-ray flux from individual clusters~\cite{Ackermann2014}, stacking analyses of large cluster samples offer a promising pathway to detect faint cumulative signals that fall below single-source detection thresholds~\cite{Huber2012,Ando2012}. Early efforts, such as a stacking of 53 clusters from the HIFLUGCS sample using Fermi-LAT data, found no significant emission~\cite{Huber2013}. However, a subsequent study of 55 HIFLUGCS clusters using Fermi-LAT data above 10 GeV detected a 4.3$\sigma$ excess, tentatively attributed to active galactic nuclei (AGN)~\cite{Prokhorov2014}. Conversely, a similar stacking analysis of 78 low-redshift clusters ($z < 0.12$) from the 2MASS survey yielded no significant signal~\cite{Griffin2014}.~\citet{Dutson2013} searched for gamma-ray emissions from 114 clusters hosting bright central galaxies (BCGs) with powerful radio sources; while a few high-significance signals ($>4\sigma$) were detected in individual objects, they could not be definitively linked to the BCGs, and the stacked analysis showed no conclusive evidence for cluster-scale emission~\cite{Dutson2013}. 
A more recent study stacking 112 clusters from the MCXC catalogue found a significant 5.8$\sigma$ detection, revealing a central point source dominated by AGN emission and a $\gamma$-ray ring coinciding with the virial shock~\cite{Reiss2018}. \rthis{The stacked signal has most recently  also been detected in cluster cores for the MCXC sample~\cite{Keshet25}.} Beyond galaxy clusters, stacking techniques have proven instrumental in probing diverse astrophysical phenomena in Fermi-LAT data, including the Extragalactic Background Light~\cite{Abdollahi2018}, star-forming galaxies~\cite{Ajello2020}, pulsars~\cite{Song23}, AGN-driven outflows~\cite{Ajello2021,Ajello2021b,McDaniel2023}, and compact sources such as FR0 galaxies~\cite{Khatiya2024}, and globular clusters~\cite{Henry2024}. Several studies have also examined potential dark matter signatures by analyzing stacked gamma-ray emissions from galaxy clusters, which leverage the high dark matter density of clusters to constrain particle physics models. A comprehensive review of these efforts can be found in ~\cite{Manna2024b}.
\\

In this work, we  extend the analysis of~\cite{Manna2024}, which investigated GeV $\gamma$-ray point-source emission from Sunyaev-Zeldovich (SZ)-selected clusters in the South Pole Telescope (SPT) catalog using 15 years of Fermi-LAT data by performing a stacking analysis. That work reported a $6.1\sigma$ detection coincident with the massive merging cluster SPT-CL~J2012$-$5649 (Abell~3667), along with six additional clusters exhibiting tentative signals ($\mathrm{TS} \geq 3\sigma$). We also looked for a similar signal from  SPT-CL~J2012$-$5649 in the DAMPE gamma-ray telescope~\cite{MannaDAMPE}. 
Here, we expand the dataset to 16.4 years of Fermi-LAT observations and focus on cumulative $\gamma$-ray emission from stacked clusters, probing fainter signals.

This manuscript is structured  as follows. Section~\ref{sec:lebel2} describes the SPT-SZ galaxy cluster sample used for the analysis.  Section~\ref{sec:level3} details the data selection criteria. The stacking methodology is presented in Section~\ref{sec:level4}, and our results are reported in Section~\ref{sec:results}. \rthis{A stacked analysis of blank fields is presented in Sect.~\ref{sec:nullresults}, whereas a similar stacked analysis of synthetic sources is discussed in Sect.~\ref{sec:simulationresults}.} We conclude with a discussion of the implications in Section~\ref{sec:conclusions}. Throughout this work, we adopt a flat $\Lambda$CDM cosmology with $\Omega_m = 0.3$ and $h = 0.7$.

\section{CLUSTER SELECTION}
\label{sec:lebel2}
The galaxy cluster sample analyzed in this work is derived from observations by SPT, a 10-meter diameter millimeter-wave telescope located at the South Pole. The SPT conducts wide-area surveys at three frequencies: 95, 150, and 220~GHz, achieving an angular resolution of $\sim\!1^\prime$ at 150~GHz~\cite{Carlstrom11}. We utilize its SZ-selected cluster catalog, which provides a nearly mass-limited sample with minimal redshift dependence, making it ideal for studying cluster populations across cosmic time. SPT completed a 2500 square degree survey from 2007 and 2011 to detect galaxy clusters. This survey was able to detect 677  galaxy clusters with  SNR greater than 4.5, corresponding to   a mass threshold of $3 \times 10^{14} M_{\odot}$ up to redshift of  1.8~\cite{Bleem15,Bocquet2019}\footnote{{\url{ https://pole.uchicago.edu/public/data/sptsz-clusters/2500d_cluster_sample_Bocquet19.fits }}}.   The  redshifts for SPT clusters were obtained using a dedicated optical and infrared follow-up campaign from targeted imaging and spectroscopic observations~\cite{Song,Ruel} in conjunction with  data from optical imaging surveys such as the Blanco Cosmology Survey~\cite{Desai12} and Dark Energy Survey~\cite{Saro}.  The SZ selection minimizes biases inherent in X-ray or optical surveys, ensuring a uniform selection function critical for statistical studies of cluster astrophysics and cosmology. 

Our analysis focuses on 300 clusters, consistent with the sample size used in~\cite{Manna2024}. The clusters are ranked in descending order of $M_{500}/z^2$, where $M_{500}$ is the total mass enclosed within a radius corresponding to 500 times the critical density of the universe at the cluster's redshift. This ranking prioritizes massive, low-redshift systems, though alternative scaling relations (e.g., $M_{500}/D_L^2$, with $D_L$ as luminosity distance) yield consistent results. 
\section{Data Selection}
\label{sec:level3}
Our analysis spans approximately 16.4 years of observations from the Fermi-LAT, covering the period from August 5, 2008 to December 8, 2024,\texttt{(MET239587201-755308805)}. We covered the energy range from 1 GeV to 300 GeV. We used 1 GeV as our lower energy threshold because the Fermi-LAT's Point Spread Function (PSF) gets degraded at lower energies. We used a spatial bin size as 0.08$^\circ$. To minimize contamination from Earth's limb, we impose a maximum zenith angle of 105$^\circ$. For each cluster source, we define a 10$^\circ$ $\times$ 10$^\circ$ region of interest (ROI) centered on its position determined by SPT observations. Standard data filters are applied, specifically \texttt{DATA\_QUAL} > 0 and \texttt{LAT\_CONFIG} == 1. The analysis leverages \texttt{FermiPy (v1.2)}, built on the \texttt{Fermitools (v2.2.0)}. Photon events are selected from the \texttt{P8R3\_SOURCE\_V3} class. To maximize sensitivity, we perform a joint likelihood analysis incorporating all four point-spread function (PSF) event types available in the Pass 8 dataset. The data are categorized into quartiles based on the quality of the reconstructed direction, ranging from the lowest quality (PSF0) to the highest quality (PSF3). Each PSF type is paired with its specific isotropic spectrum, labeled as \texttt{P8R3\_SOURCE\_V3\_PSF\textit{i}\_v1} (where \textit{i} = 0–3)~\cite{Khatiya2024}. The Galactic diffuse emission is modeled using the \texttt{gll\_iem\_v07.fits}. These models were specifically chosen to account for all the diffuse background radiation in our observations. We modeled the point source emission using the 4FGL-DR4 catalog (\texttt{gll\_psc\_v24})~\cite{Abdollahi2020}.
To account for photon leakage from sources outside the ROI caused by the detector's PSF, the model incorporates all 4FGL-DR4 sources within a 15$^\circ$ $\times$ 15$^\circ$ region. An energy dispersion correction \texttt{edisp\_bins = -1} is applied to all sources except for the isotropic component.
\textbf{We have used the publicly available code by the Fermi LAT collaboration for our analysis, available at \url{https://github.com/ckarwin/Fermi_Stacking_Analysis}}

\section{Stacking Analysis}
\label{sec:level4}
In our previous study~\cite{Manna2024}, we analyzed 300 galaxy clusters and identified seven clusters with strong $\gamma$-ray detections with significance $>3\sigma$. 
These seven clusters are SPT-CL J2012-5649, SPT-CL J2021-5257, SPT-CL J0217-5245, SPT-CL J2140-5727, SPT-CL J0232-5257, SPT-CL J0124-4301,  and SPT-CL J0619-5802. To avoid biasing the stacking analysis by including these high-significance detections, we exclude them and focus on the remaining 293 clusters with $TS  < 9.0 $. 
We calculate the test statistic (TS) for each source in the cluster catalog, defined as~\cite{Mattox}
\begin{equation}
    \mathrm{TS} = -2 \ln\left(\frac{L_0}{L}\right),
\end{equation}
where $L_0$ is the maximum likelihood under the null hypothesis (source excluded from the model) and $L$ is the maximum likelihood under the alternative hypothesis (source included). The TS quantifies the improvement in the fit when the source is added. 
Under the null hypothesis, TS follows a $\chi^2$ distribution for 1 degree of freedom (dof) according to Wilks' theorem~\cite{Wilks1938}. The detection significance is given by $\sqrt{TS}$. 
During the likelihood fitting process, we allow the normalization and spectral index of the Galactic diffuse emission to vary freely. Additionally, the normalization of the isotropic background and the normalizations of sources within a 5$^\circ$ radius of the ROI that have $TS \geq 25.0$ are treated as free parameters. For the cluster sources under analysis, we employ a power-law spectral model, leaving both the normalization and spectral index as free parameters. The power-law photon flux model is defined using the \texttt{PowerLaw2} formulation from the Fermi-LAT analysis tools~\footnote{{\url{https://fermi.gsfc.nasa.gov/ssc/data/analysis/scitools/source_models.html}}}: 
\begin{equation}  
  \frac{dN}{dE} = \frac{N(\Gamma + 1)E^{\Gamma}}{E_{\max}^{\Gamma+1} - E_{\min}^{\Gamma+1}}, 
\end{equation}
where $N$ is the total integrated photon flux over the energy range $[E_{\min}, E_{\max}]$ and $\Gamma$ is the photon index, describing the spectral slope. This parametrization ensures that \(N\) represents the total photon flux within the energy bounds, consistent with the \texttt{PowerLaw2} model in Fermi-LAT analyses. Unlike the standard \texttt{PowerLaw} model, which parameterizes flux normalization at a fixed pivot energy $E_0$, the \texttt{PowerLaw2} model directly parameterizes the integrated flux $N$ over $[E_{\min}, E_{\max}]$. This allows the likelihood framework to compute errors on $N$ directly during fitting, eliminating the need for error propagation from a pivot energy~\cite{Abdollahi2020}. To identify potential new sources within the ROI, we utilize the \texttt{find\_sources()} method from the \texttt{Fermipy} package~\cite{Wood2017}, which searches for point sources exceeding a significance threshold of $TS \geq 25.0$. The minimum angular separation between two sources is set to 0.5$^\circ$ to ensure distinct source identification. However, we do not detect any new sources meeting the criterion in our analysis.
Next, we construct two-dimensional TS profiles for each galaxy cluster. This is achieved by systematically varying two parameters for each galaxy cluster, viz.  the total integrated photon flux and the photon index. The photon flux is varied over a range of $10^{-13}$ to $10^{-9}~\mathrm{ph~cm^{-2}~s^{-1}}$, using 10 logarithmically spaced bins. Simultaneously, the photon index is varied from -4.0 to -1.0, in increments of 0.1. This produces a grid of TS values for each cluster. During the scan, the spectral parameters of all background point sources are held fixed, while the normalization of the isotropic background and both the spectral index and normalization of the Galactic diffuse emission are allowed to vary. This ensures that the model accounts for uncertainties in the background components while optimizing the fit. The scan is performed for all event classes, which are added to get a single TS profile for each source. Finally, the TS profile of each source is added to generate the final stacked TS profile. Our stacked profile is computed using the full sample of sources, ensuring an unbiased representation of the dataset. This approach has been validated in previous studies~\cite{Khatiya2024,Ajello2021}.
\section{Results}
\label{sec:results}
\begin{figure}
\hfill
\centering
\begin{adjustbox}{right=0.7\columnwidth}
\includegraphics[width=0.9\columnwidth]{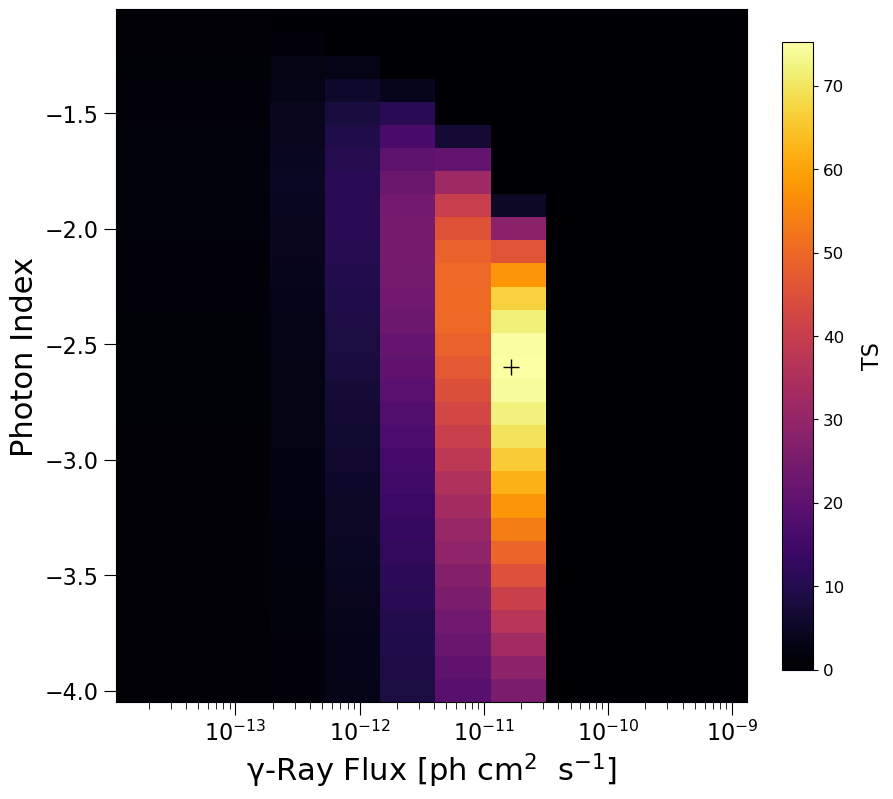}
\end{adjustbox}
\caption{Stacked $\gamma$ ray TS map of galaxy clusters with individual $TS < 9.0$. The color bar indicates the TS values, and the + sign shows the maximum TS value of 75.20. The corresponding best fit Index value is $-2.59 \pm 0.20$ and the flux is $1.67^{+1.35}_{-1.07} \times 10^{-11}$$~\mathrm{ph~cm^{-2}~s^{-1}}$. We used $0.08^{\circ}$ pixel resolution for the spatial binning.}
\label{fig:figure1}
\end{figure}

\begin{figure}
\hfill
\centering
\begin{adjustbox}{right=0.7\columnwidth}
\includegraphics[width=0.9\columnwidth]{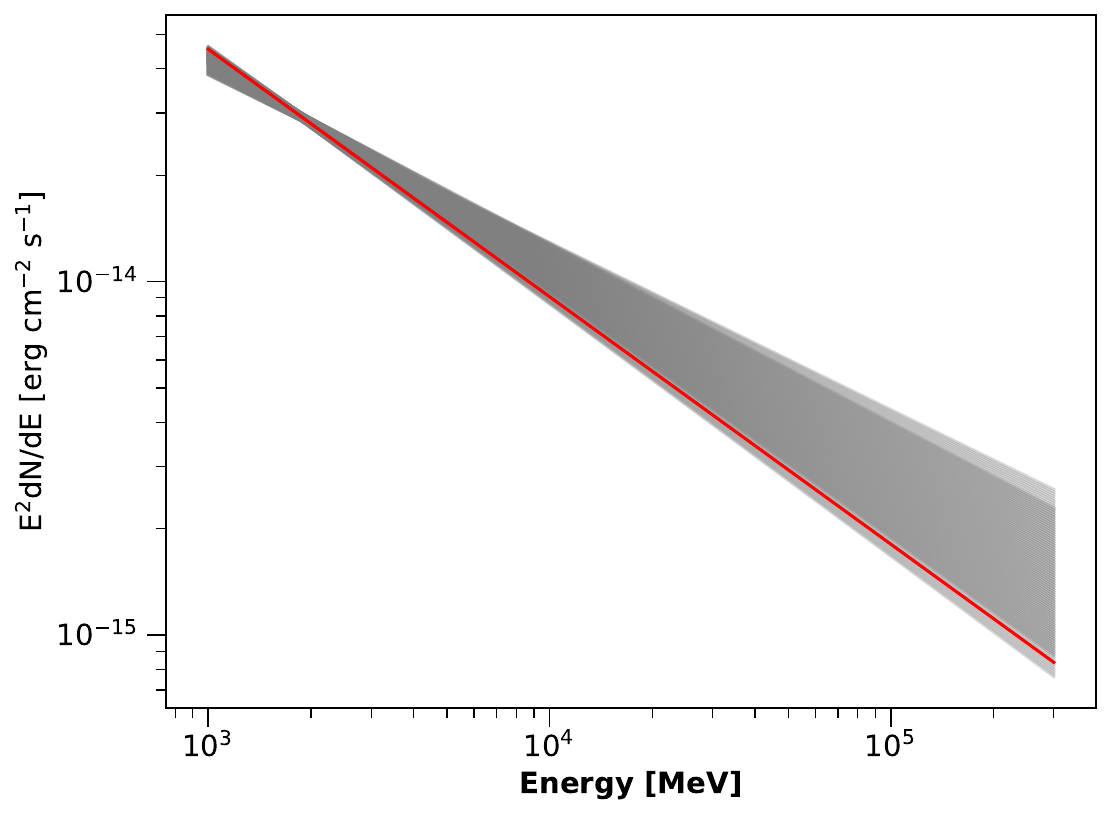}
\end{adjustbox}
\caption{Spectral Energy Distribution (SED) of the stacked signal, illustrated by the butterfly plot. The grey band corresponds to the 1$\sigma$ statistical uncertainty, and the solid red line represents the best-fit spectral model.}
\label{fig:figure2}
\end{figure}

\begin{figure}
\hfill
\centering
\begin{adjustbox}{right=0.7\columnwidth}
\includegraphics[width=0.9\columnwidth]{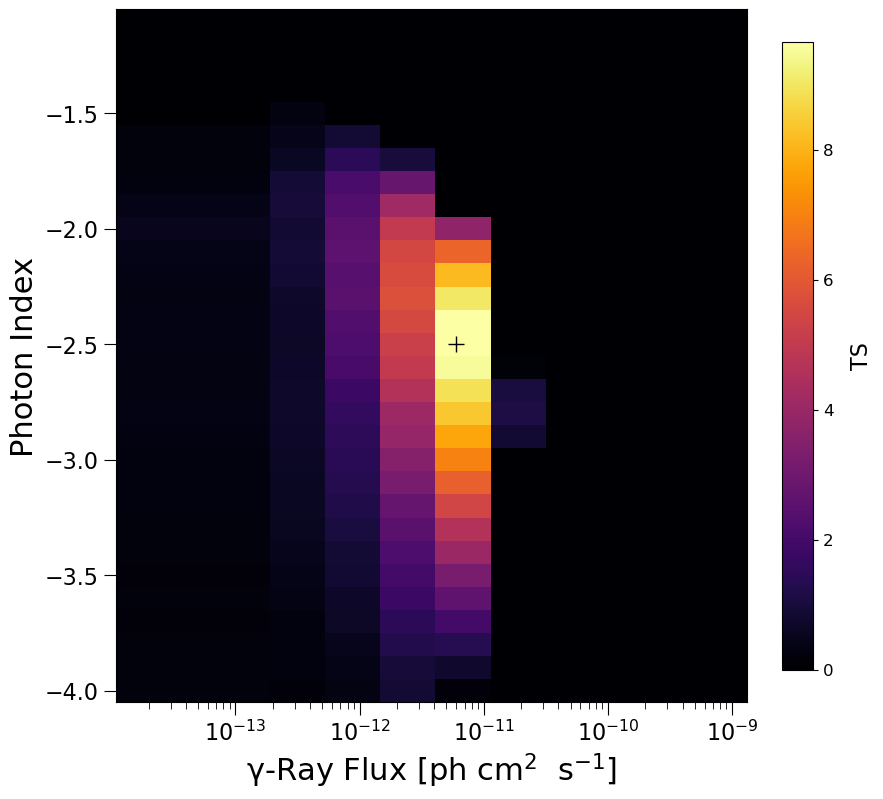}
\end{adjustbox}
\caption{Similar to ~\ref{fig:figure1} but here we depict stacked TS profile for clusters with $TS < 4.0$. The maximum TS value is 9.65. The corresponding best fit Index value is $-2.59^{+0.40}_{-0.50}$ and the flux is $6.00^{+1.07}_{-1.84} \times 10^{-12}$$~\mathrm{ph~cm^{-2}~s^{-1}}$.}
\label{fig:figure3}
\end{figure}

\begin{figure}
\hfill
\centering
\begin{adjustbox}{right=0.7\columnwidth}
\includegraphics[width=0.9\columnwidth]{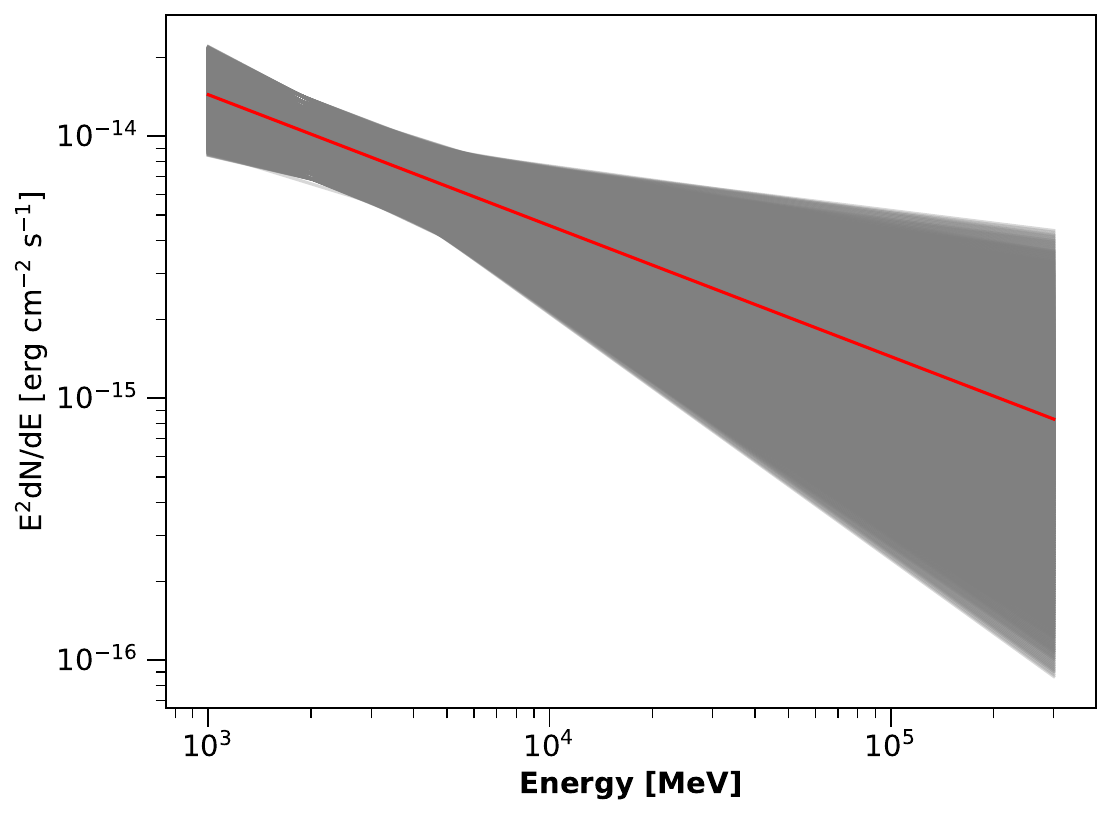}
\end{adjustbox}
\caption{Similar to~\ref{fig:figure2}, but here we depict SED (butterfly plot) for clusters with $TS \leq 4.0$.}
\label{fig:figure4}
\hfill
\centering
\begin{adjustbox}{right=0.7\columnwidth}
\includegraphics[width=0.9\columnwidth]{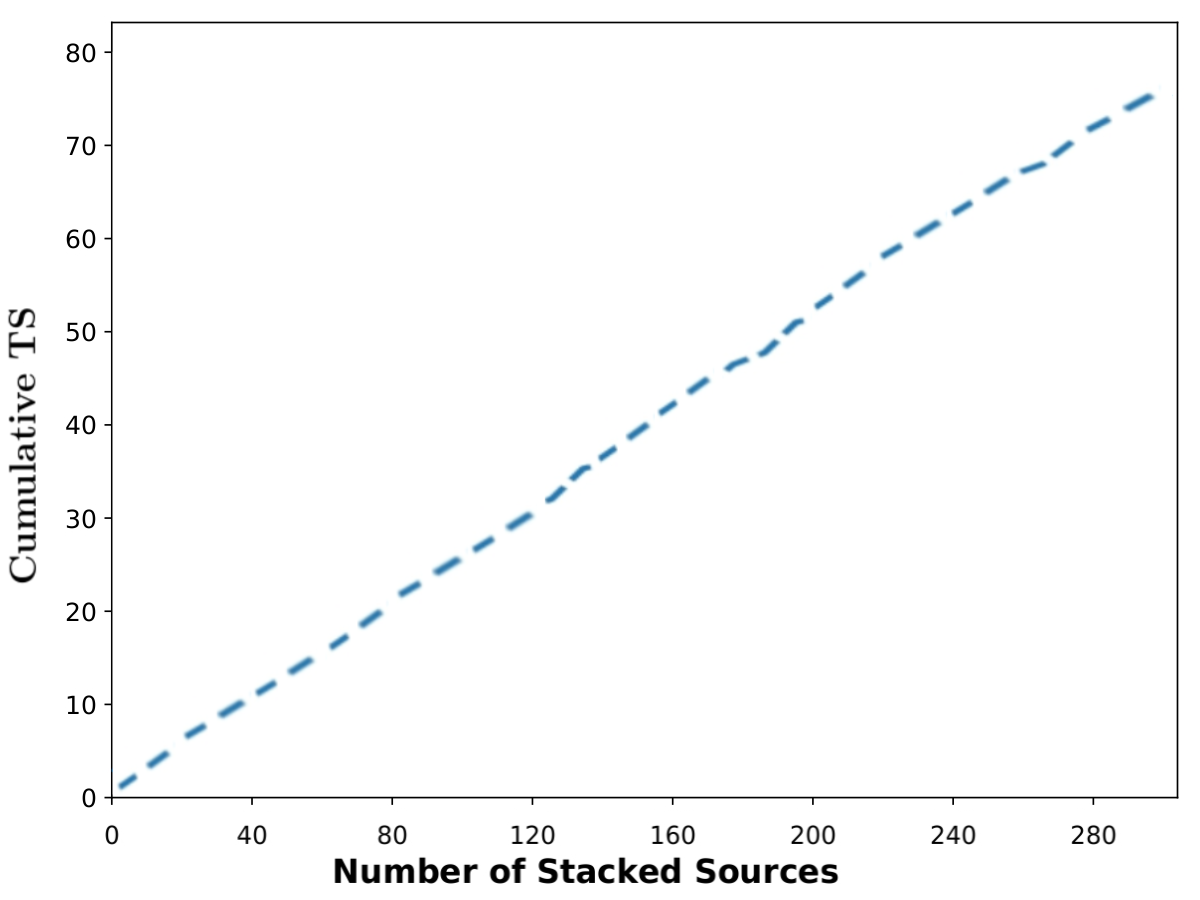}
\end{adjustbox}
\caption{\rthis{Cumulative TS as a function of the number of stacked galaxy clusters (N = 293) with individual TS values corresponding to a significance below $3\sigma$. The cumulative TS increases  linearly with the number of sources stacked.}}
\label{fig:figure5}
\end{figure}

The stacked TS profile for galaxy clusters with individual $\mathrm{TS} < 9.0$ is shown in Figure~\ref{fig:figure1}. This analysis yields a maximum TS value of $75.2$, corresponding to a detection significance of $8.39\sigma$ for two degrees of freedom, which are photon index and flux. The best-fit spectral parameters for the stacked signal are a photon index of $\Gamma = -2.59 \pm 0.20$ and an integrated flux of $1.67^{+1.35}_{-1.07} \times 10^{-11}~\mathrm{ph~cm^{-2}~s^{-1}}$. The spectral energy distribution (SED) of the stacked signal, illustrated by the butterfly plot in Figure~\ref{fig:figure2}, confirms the consistency of the power-law model across the energy range. To assess the impact of low-significance clusters, we repeated the analysis after excluding all sources with $\mathrm{TS} \geq 4.0$. \textbf{There are 123 such sources.} \rthis{The resulting TS profile for the subsample of 170 clusters ($\mathrm{TS} < 4.0$) is, shown in Figure~\ref{fig:figure3}, exhibits a weaker cumulative signal with a maximum TS of $9.65$ ($2.65\sigma$ significance)}. The best-fit photon index and flux for this subsample are $\Gamma = -2.50^{+0.40}_{-0.50}$ and $6.00^{+1.07}_{-1.84} \times 10^{-12}~\mathrm{ph~cm^{-2}~s^{-1}}$, respectively. The corresponding SED butterfly plot is presented in Figure~\ref{fig:figure4}.
We further derived 95\% confidence upper limits on the flux from the two-dimensional  TS profiles for the subsample with $\mathrm{TS} < 4.0$, as no significant detections were obtained. The resulting upper limit is $7.70 \times 10^{-12}~\mathrm{ph~cm^{-2}~s^{-1}}$.
In Figure~\ref{fig:figure5}, we present the \rthis{cumulative} TS as a function of the number of stacked sources, including only those clusters  with $TS < 9.0 $. The observed increasing trend in TS suggests a cumulative effect, where stacking additional sources enhances the overall signal.
\rthis{To further investigate the origin of the gamma-ray emission, we performed stacking analyses on our cluster sample, ordered by $M_{500}/z^2$, and explored the variation of the TS with mass. Initially, we stacked signals from 170 clusters within the 0 to 2$\sigma$ significance range. This analysis revealed that more massive clusters dominated the signal, with a maximum TS of 9.65, followed by a plateau, as depicted in Figure~\ref{fig:figure6}. This sharp initial rise could be attributed to proton-proton (pp) interactions, which are expected to produce a hadronic gamma-ray signal proportional to mass raised to the 2/3 power ($M^{2/3}$)~\cite{Huber2013}. 
Another support for the same is that the mass-dependent hadronic signal aligns with the observed dominance in the early stacking phase for the first few clusters~\cite{Huber2013}.
Extending the stacking to include all 293 clusters within the 0 to 3$\sigma$ range, we observed a more linear trend in TS, reaching a maximum of 75.2 as shown in Figure~\ref{fig:figure7}. This trend suggests a significant contribution from active galactic nuclei (AGNs) within the most massive clusters, particularly those exhibiting strong central excesses (1.5–3$\sigma$). Such clusters are likely to host AGNs, which could drive the linear signal growth. In our future works we shall explore this further.}

\section{Null Results from Blank Field Stacking}
\label{sec:nullresults}
\rthis{To assess the robustness of the gamma-ray signal detected from galaxy clusters and to exclude the possibility of stacking diffuse background emission rather than point source radiation, we conducted a null test by stacking 300 randomly selected empty sky positions. The RA range is from $1.66^\circ$ to $359.89^\circ$ and Dec. range is $-82.82^\circ$ to $87.33^\circ$. These positions were chosen to lie outside the 95\% error radii of any source in the Fermi 4FGL-DR4 catalog~\cite{DR4}, ensuring no contamination from known gamma-ray emitters. The same procedure that was used for the cluster stacking was applied to these blank fields. The resulting TS stacked map shown in Figure~\ref{fig:figure8} reveals a predominantly low-significance distribution, indicative of no significant gamma-ray detection. In contrast, our cluster stacking yielded TS values up to 8.4$\sigma$, highlighting a clear segregation between the cluster detections and  null results obtained from random sky positions.
These findings reaffirm that the gamma-ray signal observed in the cluster sample is unlikely to arise from stacking of diffuse background emission. }

\section{Stacking Test with Simulated Sources}
\label{sec:simulationresults}
\rthis{To validate our stacking methodology, we simulated 196 months of Fermi-LAT data for 100 mock $\gamma$ ray sources with power-law spectra for livetime of \texttt{(MET239587201-755308805)} which is 515721604 seconds. The photon indices were drawn from a normal distribution ( $\Gamma \sim \mathcal{N} (2.6, 0.2) $), and fluxes from a log-normal distribution with a peak at $10^{-11}~\mathrm{ph~cm^{-2}~s^{-1}}$ and a log-space dispersion of 0.2. The sources were randomly distributed in the sky away from bright Galactic regions to minimize  systematics in the background diffuse emission from near the galactic plane. We used the standard Fermi LAT tool for simulation \texttt{gtobssim}\footnote{See \url{https://raw.githubusercontent.com/fermi-lat/fermitools-fhelp/master/fhelp_files/gtobssim.txt}}. We then applied the same stacking pipeline as used earlier. The stacking analysis yielded a maximum TS of 65.5 ($7.80 \sigma$), as shown in Figure~\ref{fig:figure9} indicating a significant detection. The best-fit spectral index and integrated flux are $\Gamma = -2.64 \pm 0.20$ and $3.15^{+1.04}_{-1.11} \times 10^{-11}~\mathrm{ph~cm^{-2}~s^{-1}}$, which are consistent with the input signal parameters. These results closely match our observed signal from 300 clusters, confirming the robustness of our stacking technique for detecting faint $\gamma$ ray emission from galaxy clusters.}

\begin{figure}
\hfill
\centering
\begin{adjustbox}{right=0.7\columnwidth}
\includegraphics[width=0.9\columnwidth]{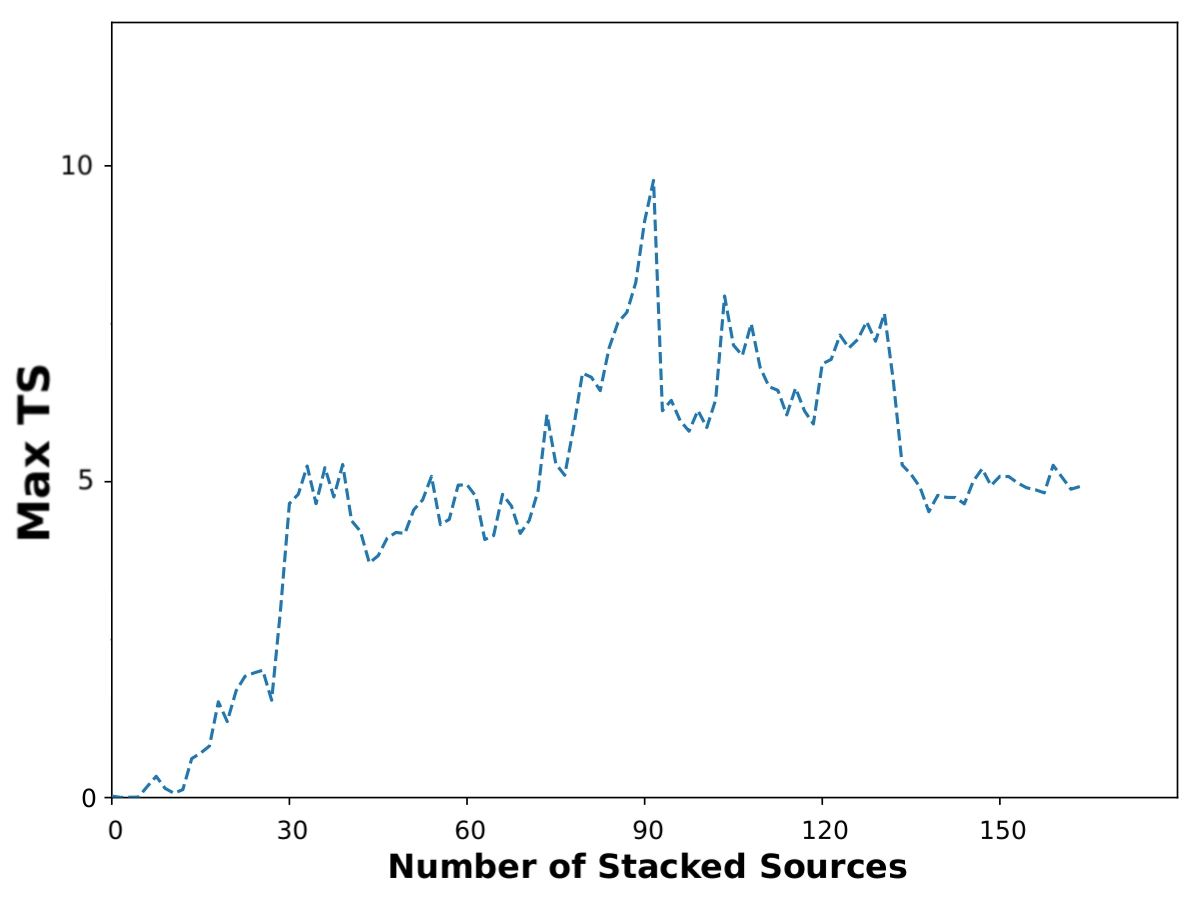}
\end{adjustbox}
\caption{\rthis{Maximum TS as a function of the number of stacked galaxy clusters (N = 170), each with an individual TS below 2$\sigma$, ordered by descending $M_{500}/z^2$. The initial rise followed by a plateau in TS suggests that the signal is primarily contributed by the most massive and nearby clusters. This behavior is consistent with a scenario in which the $\gamma$-ray emission arises from proton-proton interactions within the intracluster medium, with a peak TS of ~9.65. 170 Clusters with $TS < 4$ were stacked.}}
\label{fig:figure6}
\end{figure}

\clearpage  

\begin{figure}
\hfill
\centering
\begin{adjustbox}{right=0.7\columnwidth}
\includegraphics[width=0.9\columnwidth]{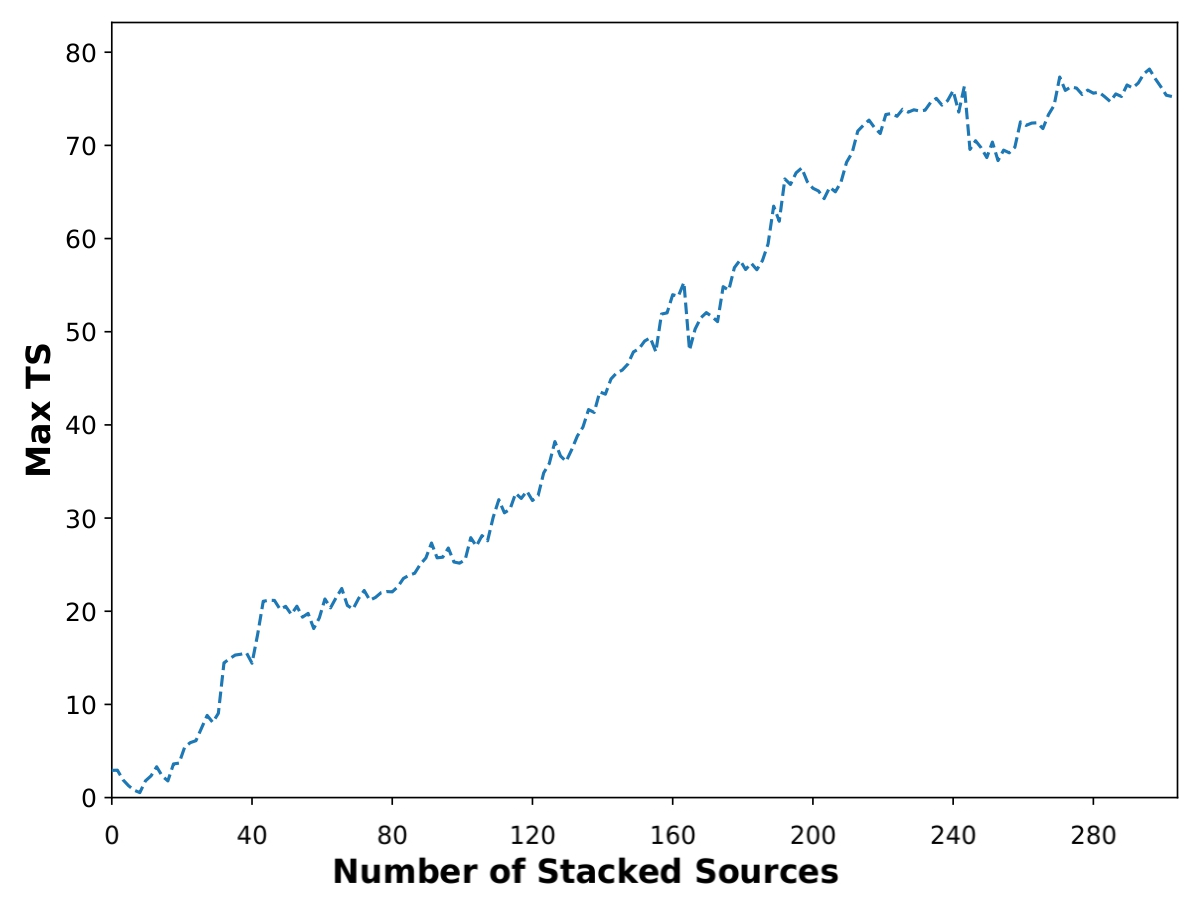}
\end{adjustbox}
\caption{\rthis{Maximum TS as a function of the number of stacked galaxy clusters (N = 293), sorted in descending order of 
$M_{500}/z^2$. The observed linear increase in maximum TS suggests that the stacking signal is dominated by a few high-TS sources, likely associated with individual bright emitters such as active galactic nuclei (AGN). All the 293 clusters with $TS < 9$ were stacked.}}
\label{fig:figure7}
\end{figure}

\begin{figure}
\hfill
\centering
\begin{adjustbox}{right=0.7\columnwidth}
\includegraphics[width=0.9\columnwidth]{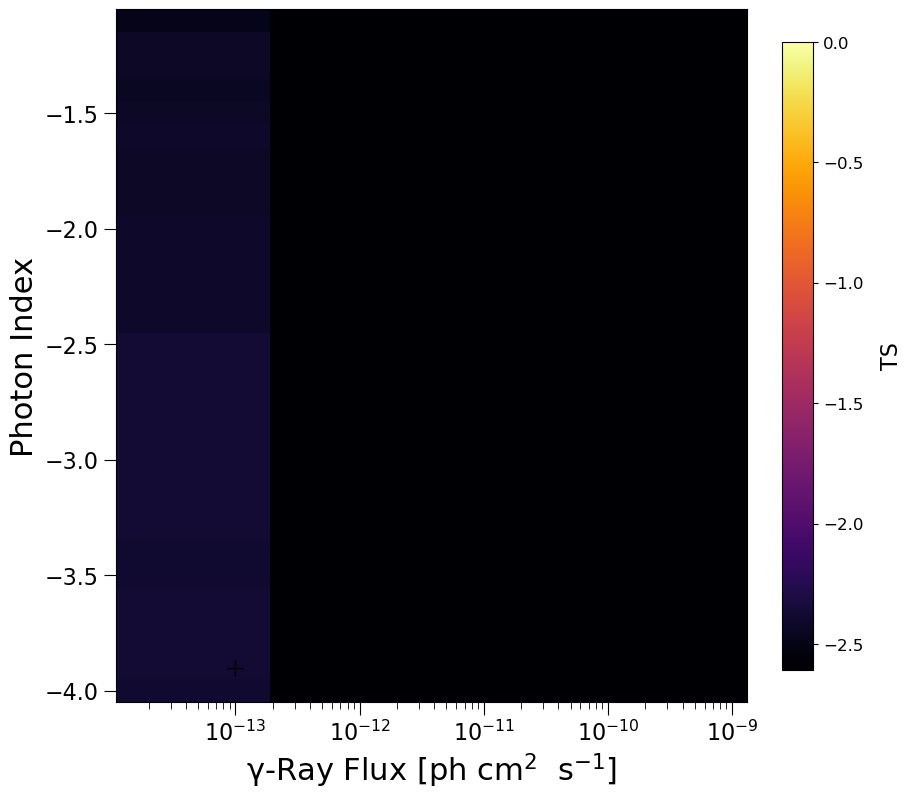}
\end{adjustbox}
\caption{\rthis{Stacked $\gamma$ ray TS map of 300 random sky positions. We do not find any statistically significant excess for this null test.}}
\label{fig:figure8}
\end{figure}
\begin{figure}
\hfill
\centering
\begin{adjustbox}{right=0.7\columnwidth}
\includegraphics[width=0.9\columnwidth]{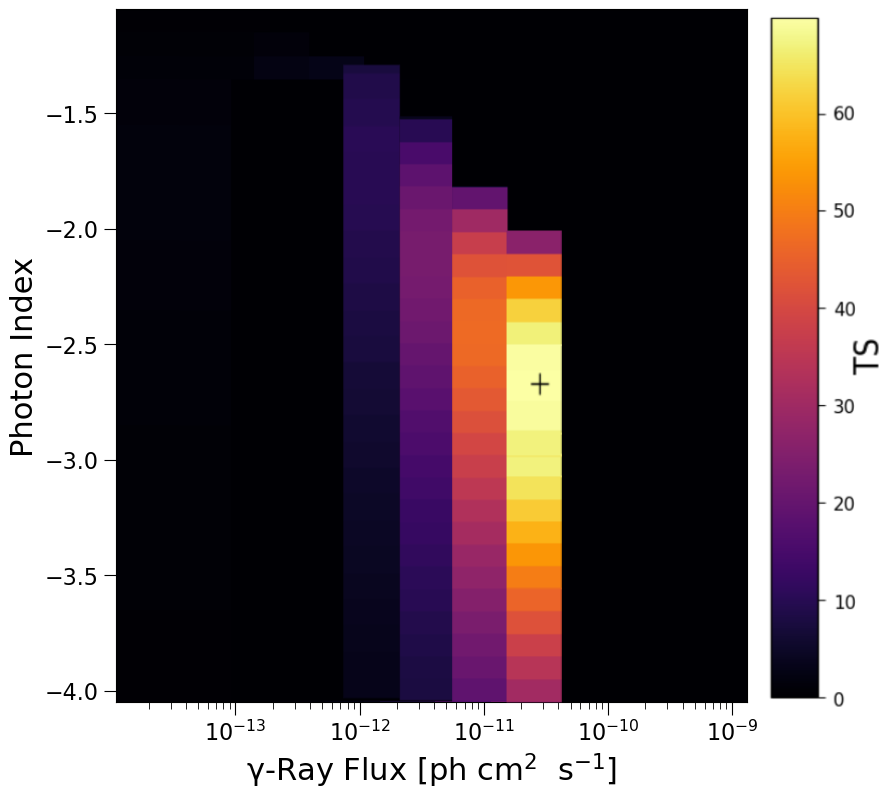}
\end{adjustbox}
\caption{\rthis{Stacked TS map for 100 simulated sources, each with a power-law gamma-ray spectrum (spectral index drawn from a Gaussian distribution centered at 2.6 with $\sigma = 0.2$) and fluxes drawn from a log-normal distribution centered at $10^{-11}$~ph~cm$^{-2}$~s$^{-1}$ ($\sigma = 0.2$ in log-space). The stacked analysis was performed over 196 months of simulated Fermi-LAT data. The TS distribution reveals a cumulative excess consistent with the input population properties, validating the stacking methodology.}}
\label{fig:figure9}
\end{figure}
\section{\label{sec:conclusions}Conclusions\protect}
We report a significant cumulative $\gamma$-ray signal from stacked galaxy clusters using 16.4 years of Fermi-LAT observations. The analysis of clusters with individual $\mathrm{TS} < 9.0$ (found in our previous analysis~\cite{Manna2024}) reveals a robust detection (TS = 75.20, $8.39\sigma$), with a best-fit power-law spectrum characterized by a spectral index of $-2.59 \pm 0.20$ and an integrated flux of $1.67^{+1.35}_{-1.07} \times 10^{-11}~\mathrm{ph~cm^{-2}~s^{-1}}$. 
In contrast, the subsample of clusters with $\mathrm{TS} < 4.0$ yields a marginal signal (TS = 9.65, $2.65\sigma$), consistent with background fluctuations, and a lower flux of $6.0^{+1.07}_{-1.84} \times 10^{-12}~\mathrm{ph~cm^{-2}~s^{-1}}$. The 95\% stacked upper limit for the subsample is found to be $7.70 \times 10^{-12}~\mathrm{ph~cm^{-2}~s^{-1}}$. These results suggest that the primary signal arises from clusters near the $\mathrm{TS} \sim 4\text{--}9$ threshold, which may harbor faint $\gamma$-ray components. 
Therefore, this detection provides compelling evidence for $\gamma$-ray emission within galaxy clusters.
Although, the  spectral index of $-2.59 \pm 0.20$ aligns with theoretical expectations for hadronic cosmic-ray interactions in the intracluster medium~\cite{Brunetti2014}, contributions from  point sources such as different types of AGNs  cannot be ruled out.

\rthis{To investigate this issue further, we then checked  the variation of  TS with  number of stacked sources,  after sorting the clusters in decreasing order of $M_{500}/z^2$. We performed separate stacking analyzes for clusters with TS $<$ 4 and those with TS $<$ 9. Among the 170 low-TS clusters (TS=9.65,$2.65\sigma$ significance), the signal appears to be dominated by hadronic interactions, as shown in Figure~\ref{fig:figure6}. When extending the stacking to include all 293 clusters (TS=75.2,$8.39\sigma$ significance), we observed a linear increase in TS with the number of stacked clusters, indicating a growing contribution from AGN-related emission as shown in Figure~\ref{fig:figure7}. When we remove the high TS values, the 
 remainder of the stacked signal could be  due to hadronic emission in the intracluster medium.}

These results highlight the power of stacking analyses in probing faint, diffuse emission from galaxy clusters and underscore the need for deeper $\gamma$-ray observations or multi-wavelength campaigns to disentangle the underlying astrophysical processes. Future studies with next-generation instruments, such as the Cherenkov Telescope Array~\cite{CTA}, will further constrain the origin of this emission and its implications for cluster astrophysics and cosmology.  

\begin{acknowledgments}
Our stacking analysis builds upon the publicly available code from the Fermi-LAT collaboration, hosted at \url{https://github.com/ckarwin/Fermi_Stacking_Analysis}. This methodology has been successfully applied in previous studies~\cite{Abdollahi2018, Ajello2020, Ajello2021, Ajello2021b, McDaniel2023, Khatiya2024}.We extend our gratitude to the Fermi-LAT stacking team for their open-source contributions and \rthis{to Jonas Sinapius for technical help in creating the synthetic dataset.} We also acknowledge the Fermi-LAT Collaboration, the Fermipy team, and the Fermi Science Tools developers for providing publicly accessible data and software, which formed the foundation of our analysis. This study was inspired by a compelling seminar delivered by Nikita Khatiya at IIT Hyderabad, which motivated us to explore this research direction. Computational work was supported by the National Supercomputing Mission (NSM), Government of India, through access to the "PARAM SEVA" facility at IIT Hyderabad. The NSM is implemented by the Centre for Development of Advanced Computing (C-DAC) with funding from the Ministry of Electronics and Information Technology (MeitY) and the Department of Science and Technology (DST). SM also extend his sincere gratitude to the Government of India, Ministry of Education (MOE) for their continuous support through the stipend, which has played a crucial role in the successful completion of our research. \rthis{We would also like to thank anonymous referee, Uri Keshet, and Franco Vazza for their useful comments and suggestions which enhanced our manuscript}. 
\end{acknowledgments}

\bibliography{references}

\end{document}